# Superconducting critical fields and anisotropy of a MgB$_2$ single crystal

G.K. Perkins*, J.Moore*, Y. Bugoslavsky*, L.F. Cohen*, J.Jun†, S.M.Kazakov†, J. Karpinski† and A.D. Caplin*

*Centre for High Temperature Superconductivity, Blackett Laboratory, Imperial College of Science, Technology and Medicine, London SW7 2BZ, UK
† Solid State Physics Laboratory, ETH, CH-8093 Zürich, Switzerland.

Despite the intense activity in the year since the discovery of superconductivity in MgB$_2$, key parameters, in particular the upper and lower critical fields $H_{c2}$ and $H_{c1}$ and their anisotropies, are not well-established, largely because of the difficulty of growing MgB$_2$ crystals. Attempts have been made to deduce these parameters from experiments on polycrystalline material,[1] but they have substantial uncertainties. $H_{c2}$ is particularly important for applications, as it is the field which quenches bulk superconductivity. In terms of understanding MgB$_2$, it is now clear that the conventional electron-phonon interaction is strong enough to account for the high transition temperature $T_c$,[2] but the consequences of the double superconducting gap[3] for the anisotropy and its dependence on temperature, are uncertain. Here we describe detailed direct measurements of $H_{c1}(T)$ and $H_{c2}(T)$ for the two principal crystallographic directions in a clean single crystal of MgB$_2$. For fields in the c-direction, $\mu_0 H_{c1}^c(0) = 0.28\pm0.01$T and $\mu_0 H_{c2}^c(0)$ is 3±0.5T; this ratio of critical fields is rather low and implies that MgB$_2$ is only just a Type II superconductor. The anisotropies of both critical fields are close to 2.

Magnetisations of high quality single crystals[4] having sharp $T_c$s of 37 to 38K were measured using a double axis Vibrating Sample Magnetometer (VSM), and with a Moving Hall Probe Magnetometer (MHPM);[5] both have a maximum field of 4 T, and the sample can be rotated with respect to the applied magnetic field and aligned to within 0.1°. We show data on one representative crystal (quantitatively similar results have been obtained on a second crystal) of approximate dimensions 500μm x 300μm x 30μm that has the c-axis normal to the platelet.

For both field orientations $H_{c2}(T)$ is identified as the field at which the reversible magnetisation vanishes (Fig. 1, inset); the advantage of the MHPM over the VSM is that the latter has a significant field-dependent background, which is almost totally absent from the former.

For the measurement of $H_{c1}$ we have used the VSM. The sample is first warmed above $T_c$ and then cooled in zero applied magnetic field. We then determine the field $H_p$ at which flux first penetrates the sample by one of two methods, depending on the temperature. At low temperatures (where at low fields flux pinning is significant, so that reversible and irreversible moments are comparable in magnitude) $H_p$ is best found[6] by measuring the onset of flux trapping after successively larger field cycles of amplitude $H_{max}$. In a plot of the remnant magnetic moment ($m_{rem}$) after each cycle as a function of $H_{max}$ (figure 2 inset), $H_p$ is readily identifiable as a sudden onset of a finite $m_{rem}$. Note that despite the presence of pinning, $H_p$ is close to the field at which the $m_{max}(H_{max})$ curve first deviates from linearity (because the reversible moment drops very sharply immediately above $H_{c1}$).

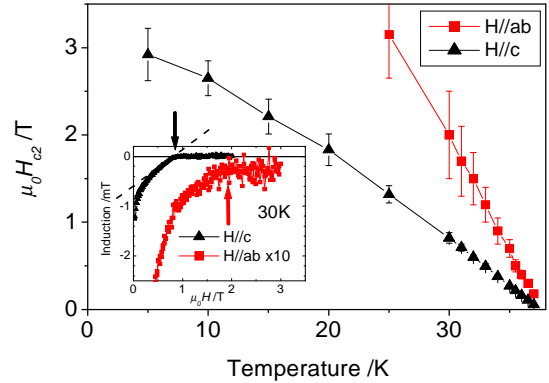

*Fig. 1. Temperature dependences of the upper critical field $H_{c2}$ for magnetic fields directed along the two principal crystallographic directions. Inset: MHPM raw data for the two principal field directions. In the MHPM, a 50 micron square Hall sensor alternates (with a period of about 2 s) in position between adjacent to the crystal, and about 300μm from it, while the applied field is held constant. The difference in Hall voltage between the positions then reflects the stray induction associated with the sample magnetisation, and $H_{c2}$ (arrowed) is identified as the applied field at which this stray induction vanishes. The very small background signal (~30μT) that is visible in the enlarged H//ab data is an instrumental artefact. The broken line indicates the slope of dM/dH for H//c as H approaches $H_{c2}$; with the MHPM data normalised so that the initial (Meissner) susceptibility matches that measured with the VSM, this dimensionless slope is 2.2±0.2 x10$^{-2}$. The slope slowly decreases to 1.0±0.3 x10$^{-2}$ at 5K. Note that the data for increasing field coincide with those for decreasing field, i.e., at these fields there is no irreversible magnetisation; furthermore, this reversibility demonstrates that there is neither supercooling nor superheating.*

At temperatures above 25K and for the H//ab orientation, $m_{rem}$ becomes too small for the onset of flux trapping to be measured reliably. However, the pinning is then very weak, so that the $m_{max}(H_{max})$ curve itself shows a sharp deviation from linearity when flux first penetrates, so giving $H_p$ directly.

For H//ab, $H_p$ can be identified with $H_{c1}$ provided that the effects of Bean-Livingston (surface) barriers[7] and geometric barriers[8] are insignificant. Geometric barriers are not expected when the field direction is in the plane of a platelet sample, and so are irrelevant for H//ab. The Bean-Livingston barrier is strongly suppressed by surface irregularities (of which our sample has many) and would not be expected to play a significant role for either field direction. Further support for these effects being small is given later.

For H//c, i.e. normal to the platelet, substantial demagnetising effects come into play. For an ellipsoid of demagnetising factor $N$, the field at the sample edge is enhanced by a factor $(1-N)^{-1}$.

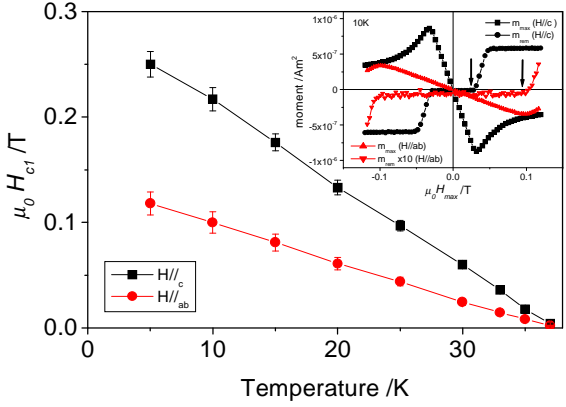

*Fig. 2. Temperature dependences of the lower critical field $H_{c1}$ for magnetic fields directed along the two principal crystallographic directions. Inset: the remanent moment $m_{rem}$ as a function of field cycle amplitude $H_{max}$, and the measured moment $m_{max}$ at the field $H_{max}$; the penetration field $H_p$ (arrowed) is that at which a finite $m_{rem}$ appears. With further increase of $H_{max}$, $m_{rem}$ increases and then saturates as expected for flux penetration into the body of the sample. The conversion of $H_p$ to $H_{c1}$ is described in the text.*

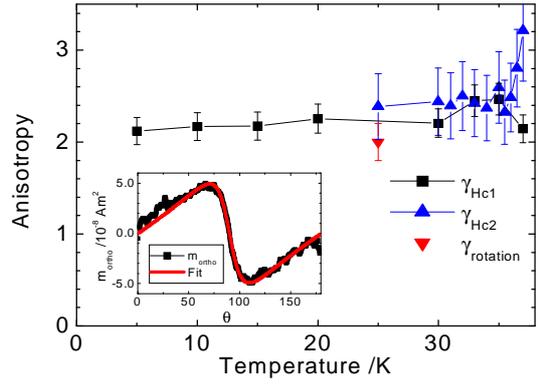

*Fig. 3 Temperature dependence of the anisotropy parameters $\gamma_{Hc1}$ ($= H_{c1}^c / H_{c1}^{ab}$) and $\gamma_{Hc2}$ ($= H_{c2}^{ab} / H_{c2}^c$). The datum point (inverted red triangle) is an independent measurement of $\gamma_{Hc1}$, obtained from measurements of the orthogonal magnetic moment as function of angle $\theta$ from the c-axis in a field of 1T (inset). The best fit of equation (1) to these data is also shown (red line).*

Our (somewhat irregular) platelet crystal, with a width $w$ to thickness $t$ aspect ratio of order 10, might be expected to behave roughly like an ellipsoid with $N \approx 1-\pi t/2w$. However, rather than relying on geometrical approximations, we use the fact that for a superconducting ellipsoid the initial magnetic susceptibility is also enhanced by the factor $(1-N)^{-1}$. We proceed from the measured low field slopes of the $m(H)$ loops in the two field directions and obtain an effective demagnetising factor $N_{eff}$ such that $(1-N_{eff})^{-1} = (dm/dH)_{H//c}/(dm/dH)_{H//ab} = 7.6 \pm 0.1$. We then equate $H_{c1}^c$ to $(1-N_{eff})^{-1} H_p$.

The anisotropy parameters $\gamma_{Hc1}$ ($= H_{c1}^c / H_{c1}^{ab}$) and $\gamma_{Hc2}$ ($= H_{c2}^{ab} / H_{c2}^c$) show no significant temperature dependences over the accessible ranges and both have values close to 2 (Fig.3). An independent evaluation of $\gamma_{Hc1}$ comes from the fluxon orientation within the crystal at intermediate magnetic fields (where surface and geometric barriers are not an issue). Due to the anisotropy, the fluxons tilt slightly away from the applied field direction towards the crystal $ab$ planes; the reversible magnetisation then has an orthogonal component $m_{ortho}$:[9]

$$m_{ortho}(\theta) \propto \ln\left(\frac{\gamma_{Hc1} \eta H_{c2}^c}{H(\sin^2\theta + \gamma_{Hc1}^2 \cos^2\theta)^{\frac{1}{2}}}\right) \frac{\sin 2\theta}{(\sin^2\theta + \gamma_{Hc1}^2 \cos^2\theta)^{\frac{1}{2}}} \quad (1)$$

where $\theta$ is the angle between the applied field and the sample c-axis, and $\eta$ is a parameter of order unity. We fit the measured $m_{ortho}(\theta)$ (Figure 3 inset) to equation (1) and obtain $\gamma_{Hc1}$. The quantity ($\eta H_{c2}^c$) enters in the logarithm only, so it has little effect on $\gamma_{Hc1}$. We obtain $\gamma_{Hc1} = 2.1 \pm 0.2$ (a factor of 10 change in ($\eta H_{c2}^c$) results in only a 10% variation in $\gamma_{Hc1}$), in excellent agreement with that obtained from direct measurement of $H_p$, strongly indicating that surface and geometric barriers have not influenced the evaluation of $H_{c1}$ from $H_p$ for either field orientation, and also that we have correctly accounted for demagnetisation factors.

For fields applied in the $c$-direction, the response is determined by the magnetic penetration depth $\lambda_{ab}$ and the superconducting coherence length $\xi_{ab}$ within the $ab$-plane. In the limit where $\kappa^c = (\lambda_{ab}/\xi_{ab}) \gg 1$, the phenomenological Ginzburg Landau (GL) theory gives

$$\mu_0 H_{c1}^c = [\Phi_0 / (2\pi \lambda_{ab}^2)] \ln(\lambda_{ab}/\xi_{ab} + 0.5) \quad (2)$$

and

$$\mu_0 H_{c2}^c = \Phi_0 / (2\pi \xi_{ab}^2) \quad (3)$$

Extrapolations to zero temperature give $\mu_0 H_{c1}^c(0) = 0.28 \pm 0.01$T and $\mu_0 H_{c2}^c(0) = 3 \pm 0.5$T. Equations (1) and (2) then yield $\lambda_{ab}(0) = 22 \pm 2$nm and $\xi_{ab}(0) = 10 \pm 0.2$ nm, hence $\kappa^c = 2.1 \pm 0.3$, violating the condition $\kappa \gg 1$. Brandt has proposed a method for calculating the reversible $M(H)$ curve for $H//c$ for any value of $\kappa > 1/\sqrt{2}$;[10] our measured ratio of $H_{c1}^c / H_{c2}^c$ of $0.08 \pm 0.005$ (over the whole temperature range) implies a value of $\kappa^c$ of $3.4 \pm 0.2$. Also, $\kappa$ can be obtained by comparing the linear slope of $M(H)$ as $H$ tends towards $H_{c2}$ (Fig. 1, inset) with Brandt's curves; this method gives $\kappa^c = 3.6 \pm 0.2$ at 30K, rising slowly to $4.3 \pm 0.2$ at 5K.

Almost all the early studies of critical fields in $MgB_2$ were on polycrystalline bulk or films, and the reported values[1] cover wide ranges. The low normal state resistivity of clean $MgB_2$[11] implies a long electronic mean free path $\Lambda$, so that in pure samples $\Lambda$ is long compared to the intrinsic superconducting coherence length $\xi_0$, hence pure material is in the "clean limit" of type II superconductivity. However when $\Lambda$ is reduced by impurity or grain boundary scattering, the effective coherence length $\xi_{eff}$ is reduced also (when $\Lambda \ll \xi_0$ an approximate expression is $\xi_{eff} = \sqrt{\xi_0 \Lambda}$). Because $H_{c2} \approx \sqrt{2} H_c \lambda / \xi_{eff}$, where $H_c$ is the thermodynamic critical field, $H_{c2}$ then increases. Thus values of $\mu_0 H_{c2}(0)$ as high as 39T have been reported in highly-disordered thin films.[12]

Only very recently have good quality $MgB_2$ single crystals become available, and even with these samples, $H_{c2}$ is best measured *via* a bulk thermodynamic property, e.g., magnetisation or heat capacity; transport studies may contain artefacts, and surface superconductivity (which can survive up to a field $H_{c3} = \sqrt{3} H_{c2}$) could be a complication.[13] The recently-reported values in single crystals of $H_{c2}^c$ as obtained from the disappearance of the magnetisation,[4,13,14] and from the heat capacity [13] are, within the experimental uncertainties, in agreement with our data shown in Figure 1, except that the crystal studied by Welp et al. [13] had a $T_c$ of only 34.5K. The temperature dependence of $H_{c2}^c$ accords with "classical" theory.[15]

Within the substantial uncertainties, our values for $H_{c2}^{ab}$ agree with those of Angst et al.[4] (on crystals from the same source as ours) obtained from the angular dependence of $H_{c2}(\theta)$, determined by the vanishing of the torque in tilted magnetic fields. At lower temperatures (where $H_{c2}^{ab}$ is above our accessible field range) they observe $\gamma_{Hc2}$ increasing with decreasing temperature, approaching a value of 6 at 15K. Zehetmayer et al.'s measurements (vanishing magnetisation)[14] of $H_{c2}^{ab}$ show a non-linear temperature dependence very close to $T_c$, and then rise rapidly at lower temperatures. Those of Welp et al. (vanishing magnetisation)[13] have a linear temperature-dependence, but one that is steeper than ours (and as already mentioned, their $T_c$ is lower). A possibility is that (as with the HTS phases) $T_c$ and $\gamma$ are sensitive to the stoichiometry. Some of the remaining differences may arise because near $H_{c2}^{ab}$ the slope of $dM/dH$ is very small (Figure 1, inset), so that its determination becomes technique-dependent.

$H_{c1}^{ab}$ in a single crystal has been measured over a very limited temperature window by Sologubenko et al.[16] using thermal conductivity; at 28K they find $H_{c1}^{ab}$ 25±2 mT, close to our value of 31±2mT. Zehetmayer et al.[14] have investigated single-crystal low field behaviour using a sensitive SQUID magnetometer; although the magnetisation loops appear similar to ours, they detect initial flux penetration at much smaller fields. The near-linear temperature dependent form of $H_{c1}^c$ and $H_{c1}^{ab}$ from our measurements is consistent with the measured temperature dependence of $\lambda$,[17] which in turn can be accounted for by the two-gap structure of $MgB_2$.[18] Finally, we note that low field torque measurements[4] yield a temperature-independent value of $\gamma$ close to 2, similar to our estimate of $\gamma_{Hc1}$.

Our values for the critical fields and their anisotropies have been obtained by direct experimental techniques, and the signatures of $H_{c1}$ and $H_{c2}$ have the form and magnitude expected for bulk transitions. While some uncertainties remain, it is clear that the $H_{c1}$ anisotropy of clean $MgB_2$ single crystals is modest, ~2, and at high temperatures $\gamma_{Hc2}$ is in the range 2 to 3; also, for $H$ parallel to the c-axis, $\kappa$ is very low. Such crystals, being in the "clean" limit, exhibit significantly different behaviour to $MgB_2$ in both bulk and thin film form; indeed, for conductor applications it will be essential to increase $\kappa$ substantially by severe reduction of the electron mean free path.

We thank Dr. M. Zehetmayer and Prof. H.W. Weber for helpful discussions. This work has been supported by the UK Engineering & Physical Sciences Research Council and by the Swiss National Science Foundation.